\documentclass{aastex}          
\usepackage{spr-astr-addons}    

\begin{document}
\graphicspath{{figures/}}
\title{Simulation of the In-flight Background for HXMT/HE}

\author{Xie Fei\altaffilmark{1}}
\and
\author{Zhang Juan\altaffilmark{1}}
\and
\author{Song Li-Ming\altaffilmark{1}}
\and
\author{Xiong Shao-Lin\altaffilmark{1}}
\and
\author{Guan Ju\altaffilmark{1}}

\altaffiltext{1}{Key Laboratory of Particle Astrophysics, Institute of High Energy Physics, Chinese Academy of Sciences, 100049 Beijing, China
xief@ihep.ac.cn}

\begin{abstract}

The Hard X-ray Modulation Telescope (HXMT) is a broad band X-ray astronomical satellite from 1 to 250 keV. Understanding the X-ray background in detail will help to achieve a good performance of the instrument. In this work, we make use of the mass modelling technique to estimate the background of High Energy Telescope (HE) aboard HXMT. It consists of three steps. First, we built a complete geometric model of HXMT. Then based on the investigation about the space environment concerning HXMT's low-earth orbit, in our simulation we considered cosmic rays, cosmic X-ray background (CXB), South Atlantic Anomaly (SAA) trapped particles, the albedo gamma and neutrons from interaction of cosmic rays with the Earth's atmosphere. Finally, the Shielding Physics List supplied by Geant4 collaborations was adopted. According to our simulation, (1) the total background of HXMT/HE is about 540 count/s on average over 20-250 keV energy band after 100 days in orbit; (2) the delayed component caused by cosmic rays and SAA trapped particles dominates the full energy band of HXMT/HE; (3) some emission lines are prominent in the background continuum spectrum and will be used for in-orbit calibration; (4) the estimated sensitivity is $\sim$ 0.1 mCrab at 50 keV with an exposure of $10^{6}$ s.

\end{abstract}


\section{Introduction}
X-ray astronomy opens an important window to investigate the universe. However, due to high atmospheric absorption, X-ray can hardly be detected on the ground. Since the X-ray telescopes have to work in a complex radiation environment above the Earth's atmosphere, the background caused by these radiation particles interacting with the sensitive detector will play an important role in the observation data, and determine the instrument sensitivity. It is necessary to analyze the radiation environment of instruments and the background level comprehensively.

The space environment consists of several components and its flux varies with the orbit. Generally, components that contribute background to X-ray instruments include but are not limited to: cosmic rays, CXB, the atmospheric albedo radiation, the radiation belts and the solar flares. The position of the instrument relative to the Earth is vital to determine the significance of each kind of component \citep{dean2003modelling}.

In a Low-Earth Orbit (LEO), the Earth's geomagnetic field can work as a natural shield against most of the low energy cosmic rays, but the flux of secondary particles induced by interaction of the primary high energy cosmic rays with the Earth's atmosphere is higher.
What's more, a special radiation region, SAA, the center of which is located at about 30S and 45W, is worth paying more attention. This area is caused by the offset of the Earth's magnetic center from geographic center, and will result in a weaker geomagnetic field at this region and much more extreme radiation environment (mainly protons and electrons) \citep{zombeck2006handbook}. Because the flux of particles is so high, in general, the instruments will switch off to protect the detector when passing through SAA. But the delayed background originating from the radioactivity of instruments illuminated by SAA may take a large proportion in the total background after passing through SAA and must be taken into account.

In a High-Earth Orbit (HEO), the SAA trapped particles have less influence on the background, because the altitude of the orbit is high above this radiation belt. Also the atmospheric albedo radiation can be ignored. However the flux of cosmic rays increases, and the solar flares contribute more to the background \citep{carter2007xmm}.

This paper is organized as follows: Section 2 shows an overview of HXMT capabilities and Section 3 gives a brief introduction of the mass modelling technology. In Section 4, we introduce the mass model of HXMT built by Geant4, which contains the geometric model of HXMT, the spectrum of the radiation environment and the physics process used in our work. In Section 5, we present the simulated results of the background. The main discussions and conclusions about this work will be presented in the last Section.

\section{The HXMT Mission}
\begin{figure}[t]
\begin{center}
  \scalebox{0.95}{ \includegraphics[width=\columnwidth]{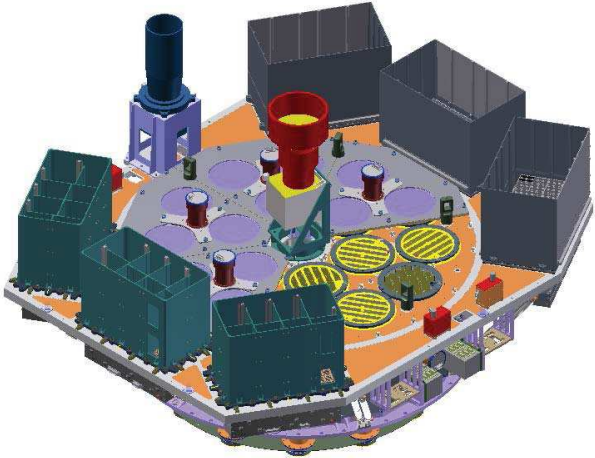}}
\caption{Structure of HXMT: The left 3 boxes are LE telescopes, the right 3 boxes are ME telescopes, and 18 HE telescopes are located in two concentric circles between the ME and LE telescopes.}
\label{hxmt1}
\end{center}
\end{figure}

HXMT works in 1-250 keV and is designed based on the direct demodulation technique \citep{li2007hxmt}. It contains three individual collimated telescopes (see Figure~\ref{hxmt1}): the High Energy X-ray Telescope (HE, 20-250  keV), the Medium Energy X-ray Telescope (ME, 5-30 keV) and the Low Energy X-ray Telescope (LE, 1-15 keV). The overview of HXMT capabilities is shown in Table~\ref{hxmtcap}.

HXMT will be launched into a 96-minute low-earth orbit with an inclination of 43$^\circ$ and an altitude of $\sim $550 km. The lifetime of HXMT is about 4 years. During this period, HXMT will mainly operate in three observation modes: all-sky imaging survey, pointed observation and deep imaging observation of some interested sky regions \citep{li2007hxmt}. The main scientific objectives of HXMT are: (1) to scan the Galactic Plane to find new transient sources and to monitor the known variable sources, and (2) to observe X-ray binaries to study the dynamics and emission mechanism in strong gravitational or magnetic fields \citep{zhang2014introduction}.

\begin{table*}
\begin{center}
\caption[]{Overview of HXMT capabilities}
\label{hxmtcap}
\begin{tabular}{cccc}
\hline
\hline\noalign{}
& HE & ME & LE  \\
\hline\noalign{}
Detector & NaI(Tl)$/$CsI(Na) & Si-PIN & SCD \\
Energy range & 20-250 keV & 5-30 keV & 1-15 keV\\
Effective area & 5000 cm$^2$ & 952 cm$^2$ & 384 cm$^2$\\
Energy resolution & $\leq$19\%(@60 keV) & $\leq$14\%(@20 keV) & $\leq$2.5\%(@6 keV)\\
Time resolution & $\leq 25\mathrm{\mu s}$ & $\leq 280\mathrm{\mu s}$ & $\leq$1ms\\
\hline\noalign{}
\end{tabular}
\end{center}
\end{table*}

\section{Mass Modelling Technique}

The method to estimate the background has been improved from the semi-empirical methods to the so-called mass modelling technique \citep{dean2003modelling}. This change benefits from the development of the computer and the skills of programming. Now the mass modelling technique, which has been used in INTEGRAL, CGRO/BATSE and many other space instruments successfully \citep{dean2003modelling}, is a mature and popular method to estimate the in-orbit X-ray background.

Actually, mass modelling technique is a physics-based method, which involves the structure of the instrument, the energy spectrum of the radiation environment and the corresponding physics processes. We implement this method by using of simulation programme, the Monte Carlo toolkit Geant4\footnote{http://geant4.web.cern.ch/geant4/}, which provides a comprehensive package including all kinds of physics processes. Under the framework of Geant4, we just need the geometrical model of telescope and the space radiation environment, and then output deposited energies in the sensitivity elements and other useful message by tracking the particles. By analyzing the output data, more information can be derived. This process can help to understand the generation of background in a completely quantitative manner.

\section{The Mass Model of HXMT}

The geometric model, input energy spectrum of the radiation environment and corresponding physics processes are three elements of mass modelling. More information about these three parts concerning HXMT will be presented in this section. The following work is based on the Geant4 Version 9.4.p04.

\subsection{The geometric model}

The geometric model contains all the information about the instrumental set-ups, including size, material, and relative position. The principles for modelling HXMT geometry are as follows:

\begin{enumerate}

\item The geometric model is close to the actual telescope as much as possible, especially for the sensitive elements and components around the sensitive elements.

\item Some complex set-ups which are far away from the sensitive detector, such as the service module, are simplified by only keeping the weight and the envelope same to the reality.

\item It is very important to describe the material composition accurately.
\end{enumerate}

\begin{figure}[t]
\begin{center}
  \includegraphics[width=\columnwidth,angle=-83]{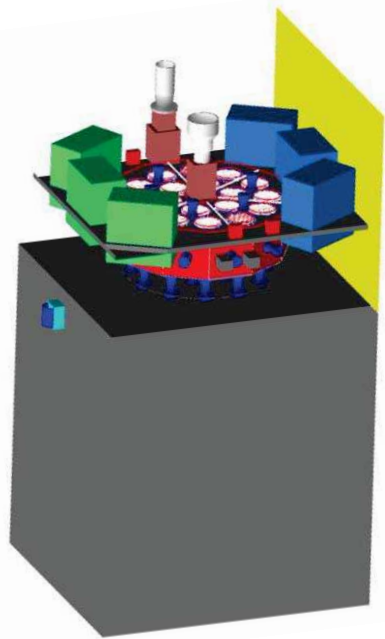}
\caption{The complete geometric model of HXMT, including HE (hidden under the anti-coincidence detectors and collimators),  ME (blue), LE (green), sun-shielding plate (yellow), the service module (grey) and the SEM (attached to the service module)
}
\label{geo_hxmt1}
\end{center}
\end{figure}

A complete and fine geometric model of HXMT has been made, which contains not only the scientific payloads (HE, ME and LE), but also the auxiliary components, like the star tracker, the space environment monitor (SEM), the sun-shielding plate and the service module. Figure~\ref{geo_hxmt1} illustrates the HXMT geometric model drawn by Geant4.

\begin{figure}[htb]
\begin{center}
  \scalebox{0.8}{ \includegraphics[width=\columnwidth]{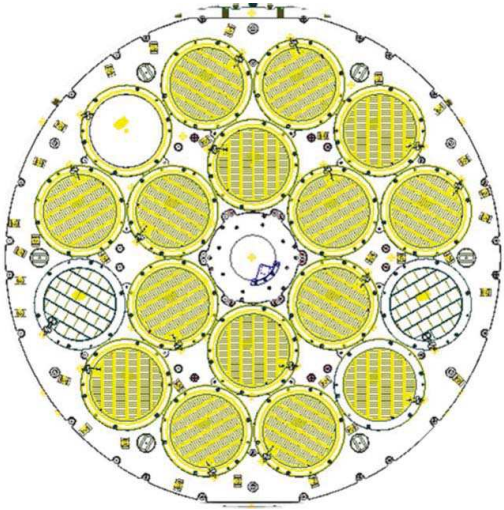}}
\caption{The details of the HE collimators: 2 large FOVs (5.71$^\circ$$\times$5.71$^\circ$), 16 small FOVs (1.14$^\circ$$\times$5.71$^\circ$), and 1 of the 16 is a blind detector which is covered by 2 mm Ta (the white one on the outer circle).}
\label{col_he1}
\end{center}
\end{figure}

\begin{figure}[htb]
\begin{center}
  \scalebox{0.8}{ \includegraphics[width=\columnwidth]{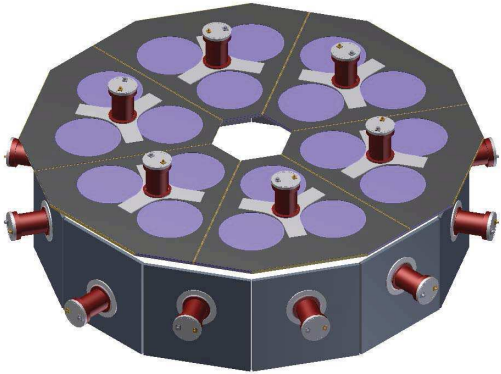}}
\caption{The layout of ACDs: 18 ACDs surround the 18 HE modules (6 ACDs on the top and 12 on the lateral side), 18 HE detectors underneath the purple circles and the red parts are photomultipliers of each ACD.}
\label{acd1}
\end{center}
\end{figure}

HE is one of three scientific payloads onboard HXMT, the simulation of which will be described in this paper. HE is composed of 18 individual detector modules, each containing a collimator (300 mm in height, made of Al and Ta), a NaI(Tl)/CsI(Na) sensitive detector (3.5 mm/40 mm) and a photomultiplier. There are two kind of FOV (Field of View) among these collimators. One is 1.14$^\circ$$\times$5.71$^\circ$, which is adopted by 16 modules. The other is 5.71$^\circ$$\times$5.71$^\circ$, 2 modules. In addition, one of the 16 small FOV collimators is covered by 2 mm Ta for measuring the local particle background. 18 collimators are divided into three groups according to the major axis directions of the FOVs and the cross angle of each group is 60$^{\circ}$ \citep{jing2010study}. Details about the collimators are shown in Figure~\ref{col_he1}. Beside the 18 detector modules, HE also has 18 Anti-Coincidence Detectors (ACD) on the top and the lateral sides. Figure~\ref{acd1} displays the structure of ACDs. The sensitive element of this veto system is a plastic scintillator with a thickness of $\sim$7 mm. It helps to identify the coincident events on NaI caused by charged particles.

%

\subsection{The spectrum of the radiation environment}
\label{sec:spect}
Since HXMT is a LEO satellite, the components of the space background that needed to be simulated are cosmic rays, CXB, SAA particles, albedo gamma and neutrons from interaction of cosmic rays with the Earth's atmosphere. Generally, the energy spectrum comes from measurements of previous experiments. Combining spectra from different experiments is a good way to obtain the input spectrum over a large energy range.

\subsubsection{CXB}

The cosmic X-ray background is generally considered to have an isotropic distribution \citep{dean2003modelling}, and its spectrum can be described as a broken power law distribution. The spectrum used in our simulation comes from Neil Gehrels \citep{gehrels1992instrumental}. In the high energy band ($>$60 MeV), the data from Fermi LAT (Large Area Telescope) observations provide some supplementary information \citep{abdo2010spectrum}.

\subsubsection{Cosmic rays}

Protons are the major component of cosmic rays in our simulation. The Earth's magnetic field has a crucial effect on the spectrum of the cosmic rays, and the minimum rigidity works as an energy cut-off in determining the incident flux. We obtain the primary cosmic-ray spectrum by fitting the measurements by AMS \citep{alcaraz2000leptons,alcaraz2000protons}, and then consider the amendment of the minimum rigidity, which is related to the geomagnetic latitude of the instrument \citep{mizuno2004cosmic}. In order to obtain a higher background level and then a conservative sensitivity estimate, only high latitude protons are simulated in this work.

\subsubsection{SAA}

SAA-induced background comes from the decay of activated isotopes. This activation process is mainly related to the high intensity cosmic proton flux in the energy range from 100 MeV to 400 MeV \citep{gang08,porras2000production}. The spectrum of protons depends on the orbit of the HXMT. The orbital period of HXMT is about 95 minutes, and the SAA passage takes about 13.3 minutes on average \citep{luphd}. Considering the procession of the orbit, HXMT will pass through the SAA averaged 12 times in a single day. Based on this information, a spectrum can be obtained simply by fitting the data from SPENVIS\footnote{https://www.spenvis.oma.be/} (ESA's Space Environment Information System).
Obviously the delayed radioactivity drops exponentially with time constants related to a series of decay processes.

\subsubsection{Albedo gamma}

The atmospheric albedo gamma radiation is caused by the interaction of the cosmic rays (mainly protons) with the atmosphere. It is generated by two mechanisms, one is the decay of mesons and the other is the bremsstrahlung of secondary electrons \citep{zombeck2006handbook}. So the flux of albedo gamma radiation strongly depends on the relative position and pointing direction of HXMT. The spectrum can be obtained by combining the measurements of atmospheric gamma rays from several instruments \citep{imhof1976high,ryan1979atmospheric}, the emissivity of the earth's atmosphere \citep{dean1989gamma} and the atmosphere model\footnote{http://ccmc.gsfc.nasa.gov/modelweb/models/nrlmsise00.php}.

\subsubsection{Albedo neutron}

Just like the albedo gamma rays, albedo neutron is also a secondary component that is caused by the interaction of the cosmic rays with the atmosphere. Inelastic scattering, elastic scattering and capture are the most possible ways for neutrons interacting with the spacecraft \citep{dean2003modelling}. The spectrum of albedo neutron was obtained by fitting COMPTEL measurements with a segmented power-law distribution \citep{armstrong1973calculations}.

\subsection{The physics processes}

Geant4 provides comprehensive physics processes. It is important to make a physics list for your application which contains all the relevant particles, physics processes and cut-off parameters. The physics processes can be simply divided into three categories: general processes (decay, transportation), electromagnetic processes (standard/low energy) and hadronic processes. Many reference physics lists are offered by the Geant4 collaboration now, and we have chosen the Shielding Physics List in our mass model (based on Geant4 Version 9.4.p04), which can be found in CERN webpage\footnote{http://indico.cern.ch/event/62629/session/7/contribution/26/
material/slides/1.pdf}.
It should be mentioned that to get more accurate simulation, we add the low-energy electromagnetic process and the radioactive decay to this shielding model.

\section{The Simulation Results}

\subsection{The background counting rates}

In order to get the background of HXMT/HE, five individual incident components (mentioned in section 4.2) are simulated. Enough events are simulated so that the relative statistical error of the background induced by each component is not larger than 1\%. According to the response time recorded by Geant4 when tracking the particles, we classify the signals into prompt ones and delayed ones. The prompt background is caused by prompt processes, such as scattering and ionization, and the delayed background comes from the decay of activated nuclei, which depends on the half-lives of the elements involved.
When the energy deposition time on a sensitive element (NaI) is shorter than 1 microsecond, it is a prompt count, and needs to be vetoed by the anti-coincidence system ACDs with threshold 100 keV. We define the counts on NaI before veto by ACDs as veto-off counts and the remaining after veto as veto-on. From the simulated data we find that the veto-on count rates are much less than the veto-off count rates for the charged particles. Taking cosmic ray protons as an example, the vote-off count rate is 440 count/s while the vote-on count rate reduces to 50 count/s.

The count rate of each component together with the statistical error is shown in Table~\ref{hxmtbg}. It also presents the total background of HXMT/HE after 100 days in orbit. According to the simulation results, 20-250 keV background is about 540 count/s on average. Considering the effective area and the full energy range, it can be translated to about 5.4$\times10^{-4}$ $count\cdot cm^{-2} \cdot s^{-1} \cdot keV^{-1}$, which is comparable to the HEXTE (The High Energy X-ray Timing Experiment) background level during in-orbit checkout phase \citep{rothschild1998flight}. Table~\ref{hxmtbg} also shows that the prompt background mainly comes from the CXB, the cosmic rays, the albedo gamma, and the delayed background comes from the cosmic rays and the SAA trapped particles. It is evident that the albedo neutron makes little contribution to the total background.

\begin{table*}
\begin{center}
\caption[]{Total background of HXMT/HE 18 modules after 100 days in orbit}
\label{hxmtbg}
\begin{tabular}{cccc}
\hline
\hline\noalign{}
Component & Averaged count rate (count/s) & Statistical error (count/s) \\
\hline\noalign{}
CXB (prompt) & 51.94 & 0.23\\
Albedo Gamma (prompt) & 86.34 & 0.25\\
Cosmic Proton (prompt/delayed) & 32.81/110.72 & 0.13/0.24\\
Albedo Neutron (prompt/delayed) & 3.32/1.48 & 0.0097/0.0065\\
SAA (delayed) & 254.01 & 1.64\\
\hline\noalign{}
Total & 540.62 & 2.51 \\
\hline\noalign{}
\end{tabular}
\end{center}
\end{table*}

\subsection{The background spectrum}

\begin{figure}[htb]
\begin{center}
  \scalebox{0.8}{ \includegraphics[width=\columnwidth]{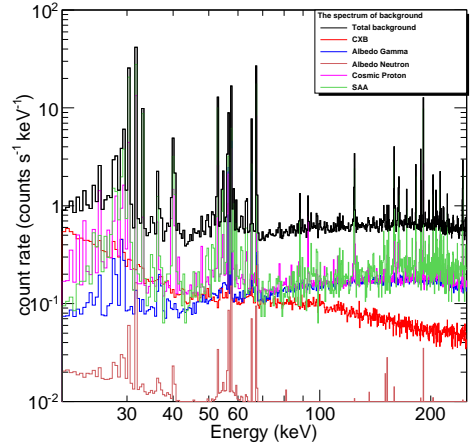}}
\caption{Simulated total background spectrum of HXMT/HE together with the individual components. The total background (dark line) consists of the contributions from CXB (red line), Albedo gamma (blue line), Albedo neutron (brown line), Cosmic proton (pink line) and SAA trapped particle (green line).}
\label{spectrum}
\end{center}
\end{figure}

Figure~\ref{spectrum} shows the background spectra induced by different components and the total one. The energy resolution and electronic noise haven't been considered in our simulation.
From the normalized veto-on spectrum, cosmic rays and SAA component dominate the background over the full energy range of HXMT/HE. It means that most of the background counts come from the delayed component, which cannot be rejected by the coincidence veto systems.


There are many emission lines superposed on the background continuum spectrum. Among all these prominent background lines, peaks at about 56 keV, 57 keV, 65 keV and 67 keV come from the fluorescence lines of tantalum, which is the main material of the collimators. Peaks due to the decay of iodine are at $\sim$ 30 keV, $\sim$ 50-60 keV, $\sim$ 60-70 keV and $\sim$ 190 keV; note that these lines were also prominent in the HEXTE background spectrum \citep{rothschild1998flight}.

\subsection{The SAA passage}

\begin{figure}[htb]
\begin{center}
  \scalebox{0.9}{ \includegraphics[width=\columnwidth]{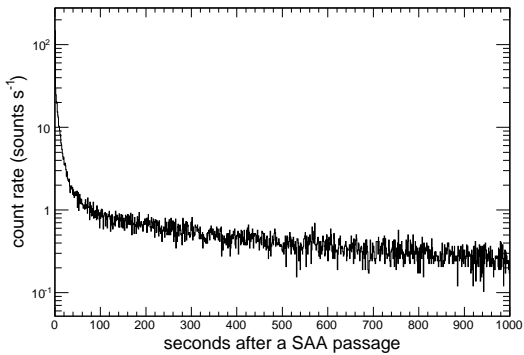}}
\caption{Illustration of the decrease with time for SAA induced background.}
\label{SAA_short}
\end{center}
\end{figure}

\begin{figure}[htb]
\begin{center}
  \scalebox{0.9}{ \includegraphics[width=\columnwidth]{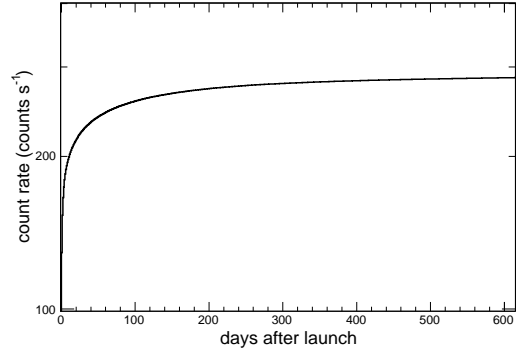}}
\caption{The long-term increase with time of SAA-induced background due to the activation of long-lived nuclides.}
\label{SAA_long}
\end{center}
\end{figure}

HXMT/HE will reduce HV (High Voltage) when passing through SAA, but the delayed count rate due to the decay of activated isotopes is the most significant contributor to total background according to Table~\ref{hxmtbg}. The subsequent decay lasts from several seconds to days which illustrates that the activated isotopes have different half-lives. Figure~\ref{SAA_short} shows the variation of count rate due to the decay of activated isotopes with time. It is clear that the induced radioactive decay is mainly concentrated in the initial several minutes, indicating that most of these activated isotopes have a short life time.
For the long-lived isotopes, their contribution is visible in the long-term accumulation effect, which is shown in Figure~\ref{SAA_long}. The SAA induced background at different periods is given, considering hundreds-of-days in-orbit operation. It is clear that SAA-induced background increases rapidly during the first $\sim$3 months due to the activation of long-lived isotopic nuclides, and gradually approaches stability after the first year's operation.

%

\subsection{Sensitivity estimation}

With the simulated background spectrum, the sensitivity which is an important parameter for a detector can be calculated.
The sensitivity estimation is given in Equation~\ref{equF} \citep{peterson1975instrumental}.

\begin{equation}
F_{min}=n_{\sigma}\cdot \frac{\sqrt{B}}{\varepsilon \cdot A_{eff} \cdot \sqrt{T_{int}} \cdot \Delta E}
\label{equF}
\end{equation}

where $F_{min}$ is the sensitivity of HXMT/HE at a given energy E with an energy band $\Delta$E of E/2; $n_{\sigma}$ is the significance; $B$ is the count rate of background in unit of count$\cdot \mathrm{s}^{-1}$; $\varepsilon$ is detection efficiency; $A_{eff}$ is the effective area, which is about 4276 $\mathrm{cm} ^{2}$ considering the shielding of the collimator; and $T_{int}$ is the exposure time.

\begin{figure}[htb]
\begin{center}
  \scalebox{0.8}{ \includegraphics[width=\columnwidth]{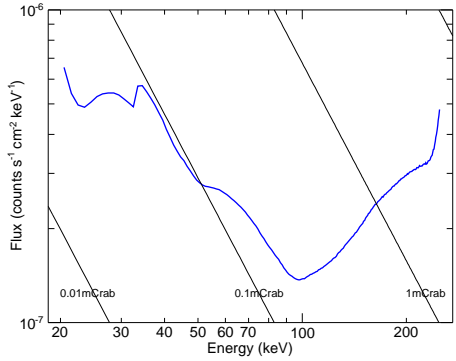}}
\caption{Simulated continuum sensitivity for HXMT/HE ($n_{\sigma}$=3, $T_{int}$=$10^{6}$ s).}
\label{sensitivity}
\end{center}
\end{figure}

Figure~\ref{sensitivity} shows the simulated continuum sensitivity for HXMT/HE with an exposure of $10^{6}$ s. The estimation of statistical sensitivity is $\sim$ 0.1 mCrab at 50 keV.

\section{Discussion and Conclusion}

The whole simulation process of HXMT/HE based on mass modelling technique has been introduced in the paper, including the geometric model of HXMT, the energy spectrum of the radiation environment and the physics processes needed, based on Geant4.9.4.p04.

We derived the background count rates induced by different components together with the total background from simulated data, which are shown in Table~\ref{hxmtbg}. The total 20-250 keV background is about 540 count/s on average after 100 days in orbit and the delayed counts due to the SAA passage is the most significant contributor, which is about 47\% of the total count rate. After each passage of SAA, the delayed background decreases rapidly in the initial several minutes due to the activation of short-lived nuclides. The accumulated count rate due to historic SAA passage increases rapidly in the first $\sim$ 100 days and approaches a steady state after about 1 year as a result of the activation of long-lived nuclides. This implies that SAA will dominate HE background count rate along the life time of HXMT.

Then from the normalized spectrum (Figure~\ref{spectrum}), we showed that the delayed component from the cosmic ray and SAA trapped particles dominates the entire energy band of HXMT/HE. The anti-coincidence shielding system ACDs can reduce the prompt background induced by charged particles effectively. Also, we showed several prominent lines in the background continuum spectrum. They come from the material of collimators and detector elements. These lines can be used for in-orbit calibration.

At last, we showed the HXMT/HE sensitivity estimation based on the simulated background level, which is $\sim$ 0.1 mCrab at 50 keV with an exposure of $10^{6}$ s. It will help us to draw up observation plans to achieve the science goals during its lifetime.

The simulated background level of HXMT/HE is obtained under an ideal assumption of the averaged incident space radiation environment. In fact, these environment components vary dramatically with time for in-flight instrument, not only in their intensity, but also the relative weight. The background caused by albedo gamma depends on the orientation of the satellite, but the difference between the maximum and the minimum is about 4 count/s, which can be ignored compared to the dominant components. Cosmic protons are restrained by the minimum rigidity of geomagnetic field. A higher proton flux corresponding to a high geomagnetic latitude region is adopted in our simulation, the prompt and delayed count rates are 32.81 and 110.72 count/s; but if a lower flux in a low geomagnetic region is used, the count rates will be 7.12 and 23.09 count/s. Therefore, the background level is overestimated for a more reliable sensitivity. Meanwhile, we expect a better sensitivity in-orbit. Results in this paper will be helpful in further study, and more works for estimation model is underway.

\acknowledgments
We thank Drs. Fan Lei and Ming Xu for their great help with Geant4 mass modelling. And we are grateful to Prof. Wei Cui for his suggestions and discussions. This work is supported by National Natural Science Foundation of China under the grant No. 11403026, by 973 Program of China under grant 2014CB845800, and by the Strategic Priority Research Program on Space Science, the Chinese Academy of Sciences under grant No. XDA04010300.

\bibliographystyle{spr-mp-nameyear-cnd}  
\bibliography{bibliography}              

\end{document}